\documentclass[conference]{IEEEtran}
\usepackage[font=footnotesize]{caption}
\IEEEoverridecommandlockouts
\usepackage{booktabs}
\usepackage{lipsum}
\usepackage{xcolor,soul,framed} 

\usepackage{diagbox}
\usepackage{array}
\usepackage{kantlipsum}
\usepackage{caption}
\usepackage{subcaption}

\usepackage{xcolor,soul,framed} 

\colorlet{shadecolor}{yellow}
\usepackage[pdftex]{graphicx}
\graphicspath{{../pdf/}{../jpeg/}}
\DeclareGraphicsExtensions{.pdf,.jpeg,.png}
\usepackage{multirow}
\usepackage[cmex10]{amsmath}
\usepackage{pgfplots}
\usepackage{tcolorbox}
\usepackage{amsfonts}  
\pgfplotsset{compat=1.17}

\usepackage{array}
\usepackage{mdwmath}
\usepackage{mdwtab}
\usepackage{eqparbox}
\usepackage{url}
\usepackage{capt-of}  
\usepackage{cuted}    
\begin{document}


\title{Enhancing Cybersecurity in Critical Infrastructure with LLM-Assisted Explainable IoT Systems\\
}

\author{
\IEEEauthorblockN{\IEEEauthorrefmark{2}Ashutosh Ghimire, \IEEEauthorrefmark{2}Ghazal Ghajari, \IEEEauthorrefmark{2}Karma Gurung,
\IEEEauthorrefmark{4}Love K. Sah,
\IEEEauthorrefmark{2}$^{*}$Fathi Amsaad\thanks{* Corresponding author}
}
\IEEEauthorblockA{
\IEEEauthorrefmark{2} Department of Computer Science and Engineering, Wright State University\\
\IEEEauthorrefmark{4} Department of Electrical and Computer Engineering, Western New England University\\
Email: \IEEEauthorrefmark{2}\{ashutosh.ghimire, ghajari.2, gurung.14, fathi.amsaad\}@wright.edu, \IEEEauthorrefmark{4} love.sah@wne.edu\\
}
}
 
\maketitle

\begin{abstract}
Ensuring the security of critical infrastructure has become increasingly vital with the proliferation of Internet of Things (IoT) systems. However, the heterogeneous nature of IoT data and the lack of human-comprehensible insights from anomaly detection models remain significant challenges. This paper presents a hybrid framework that combines numerical anomaly detection using Autoencoders with Large Language Models (LLMs) for enhanced preprocessing and interpretability. Two preprocessing approaches are implemented: a traditional method utilizing Principal Component Analysis (PCA) to reduce dimensionality and an LLM-assisted method where GPT-4 dynamically recommends feature selection, transformation, and encoding strategies.

Experimental results on the KDDCup99 10\% corrected dataset demonstrate that the LLM-assisted preprocessing pipeline significantly improves anomaly detection performance. The macro-average F1 score increased from 0.49 in the traditional PCA-based approach to 0.98 with LLM-driven insights. Additionally, the LLM generates natural language explanations for detected anomalies, providing contextual insights into their causes and implications. This framework highlights the synergy between numerical AI models and LLMs, delivering an accurate, interpretable, and efficient solution for IoT cybersecurity in critical infrastructure.
\end{abstract}



%
\IEEEpeerreviewmaketitle

\section{Introduction}

Critical infrastructure, including power grids, healthcare systems, and transportation networks, relies heavily on Internet of Things (IoT) devices for automation, monitoring, and operational efficiency. However, this integration introduces substantial cybersecurity risks, as IoT systems generate heterogeneous, high-volume data and operate on resource-constrained devices \cite{djenna2021internet}. Traditional security mechanisms often fail to manage these complexities, leaving critical systems vulnerable to threats such as Denial-of-Service (DoS) attacks, network intrusions, and data breaches \cite{aslan2023comprehensive}.

Anomaly detection methods for IoT systems predominantly utilize machine learning techniques, including clustering, Support Vector Machines (SVMs), and deep learning models like Autoencoders. While effective at identifying anomalies, these approaches face three significant challenges: (1) operating as opaque "black box" models, which lack interpretability and fail to provide explanations for their decisions, (2) neglecting robust preprocessing despite the noisy and redundant nature of IoT data, and (3) underutilizing valuable unstructured textual data, such as logs or event descriptions, which hold critical contextual insights \cite{chandola2009anomaly}.

To address these challenges, a hybrid framework is introduced that integrates Large Language Models (LLMs), such as GPT-4, with Autoencoder-based anomaly detection for IoT cybersecurity. The framework enhances anomaly detection performance through optimized, LLM-assisted preprocessing and improves interpretability by generating human-readable explanations for detected anomalies \cite{su2024large}. Specifically, two preprocessing pipelines are employed: a traditional PCA-based approach for dimensionality reduction and an LLM-driven approach, where GPT-4 recommends feature selection, transformation, and encoding strategies.

The contributions of this work are twofold. First, the proposed LLM-assisted preprocessing pipeline intelligently handles noisy, redundant, and categorical features in IoT data, leading to a significant improvement in anomaly detection performance, as demonstrated by an increase in the macro-average F1 score from 0.49 (PCA-based preprocessing) to 0.98 (LLM-assisted preprocessing). Second, GPT-4 enhances the explainability of the framework by producing natural language descriptions of anomalies, enabling cybersecurity analysts to understand and trust the model's outputs.By addressing these limitations, the proposed framework demonstrates the synergy between numerical AI models and LLMs, achieving robust, interpretable, and efficient anomaly detection. This work not only improves detection accuracy but also fosters trust in AI-driven IoT cybersecurity systems through intelligent preprocessing and explainable results.

\begin{figure*}[!t]
    \centering
    \includegraphics[width=0.9\textwidth]{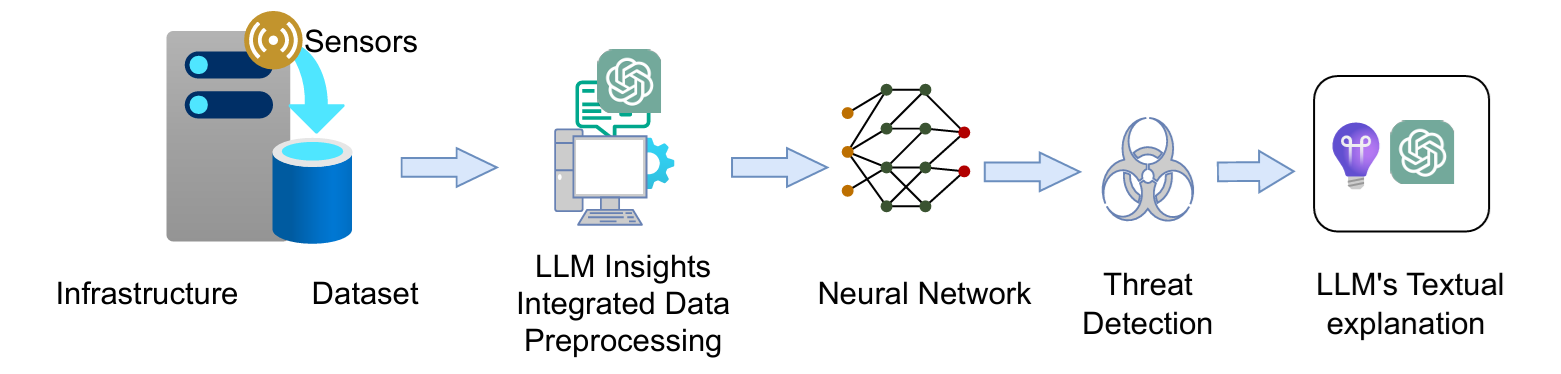}
    \caption{Proposed framework for LLM-assisted anomaly detection in IoT systems.}
    \label{fig:framework}
\end{figure*}

\section{Related Works}
The integration of artificial intelligence (AI) and machine learning (ML) techniques into IoT cybersecurity has been a topic of extensive research. Given the increasing complexity of IoT networks and their critical role in infrastructure systems, various approaches have been explored to detect, mitigate, and explain cyber threats \cite{liu2019secure}. This section reviews existing methods, highlights their limitations, and identifies the gap that this research aims to address.
 \subsection{AI and ML-Based Anomaly Detection in IoT Cybersecurity}
 In recent years, traditional and deep learning techniques have been widely applied to anomaly detection in IoT systems. Supervised learning methods such as Support Vector Machines (SVMs), Decision Trees, and Random Forests have been employed to classify IoT network traffic into normal and attack categories \cite{ghimire2023implementation}. For instance, datasets like KDDCup99 and NSL-KDD have been used to train classification models for intrusion detection. While these models are effective when labeled data is available, their reliance on labeled training samples limits their applicability to real-world IoT systems, where data labeling is time-consuming and expensive\cite{ravipati2019intrusion}.

To overcome the limitations of supervised learning, unsupervised learning methods such as clustering (e.g., k-Means, DBSCAN) and anomaly detection techniques have been explored. These methods identify patterns or deviations in unlabeled data that signify anomalies. Among these approaches, Autoencoders have shown significant promise. Autoencoders are deep neural networks that learn to reconstruct input data and flag instances with high reconstruction error as anomalies. Researchers have applied Autoencoders to IoT traffic analysis, time-series data, and sensor readings to detect abnormal behaviors indicative of cyberattacks \cite{gao2022tsmae}. However, despite their effectiveness in identifying anomalies, Autoencoders suffer from a lack of interpretability. They function as black-box models, providing no meaningful explanation for why a particular instance was flagged as an anomaly.


\subsection{The Role of Natural Language Processing and Large Language Models in Cybersecurity}
Natural Language Processing (NLP) has emerged as a promising tool for addressing the limitations of traditional numerical AI models in cybersecurity. While traditional anomaly detection focuses primarily on structured, numerical IoT data, real-world IoT systems generate substantial amounts of unstructured textual data, including logs, alerts, and error messages \cite{bertero2017experience}. NLP techniques can extract meaningful insights from this textual data, enhancing the context and interpretability of anomaly detection systems.

Recently, Large Language Models (LLMs), such as OpenAI's GPT-4, have demonstrated unprecedented capabilities in understanding and generating human-like text. In the context of cybersecurity, LLMs have been applied to various tasks, including log analysis, where they summarize system logs, detect anomalies in textual alerts, and identify patterns in error messages \cite{hadadi2024anomaly}. Additionally, LLMs are used for incident reporting, generating natural language explanations of detected incidents to assist human operators, and for threat intelligence, which involves extracting knowledge from security reports and describing potential attack vectors \cite{o2023deployment}.

Despite significant advancements, Large Language Models (LLMs) have not been effectively integrated with numerical anomaly detection models. Most existing approaches treat numerical and textual analyses as separate tasks, failing to combine their strengths to create a holistic cybersecurity solution. IoT data is inherently noisy, redundant, and heterogeneous \cite{wang2017heterogeneous}. Existing numerical AI models often depend on manual or simplistic preprocessing methods, which limit their performance. While LLMs offer enhanced interpretability, they remain underutilized in preprocessing IoT data—a critical bottleneck in anomaly detection pipelines.


\section{Proposed Framework}
This section outlines our proposed framework, which integrates numerical anomaly detection models with Large Language Models (LLMs) to address limitations in traditional IoT cybersecurity solutions. The framework consists of three main components: data preparation, anomaly detection, and natural language-based anomaly explanation. Two distinct approaches are implemented for data preprocessing: (1) PCA-based numerical preprocessing and (2) LLM-assisted preprocessing.

\subsection{Dataset}
The experiments in this study are conducted on the KDDCup99 10\% corrected dataset, a standard benchmark for intrusion detection systems \cite{tavallaee2009detailed}. The dataset comprises numerical features (e.g., \textit{src\_bytes, dst\_bytes}) and categorical features (e.g., \textit{protocol\_type, flag}). Despite its utility, the dataset poses challenges such as redundancy, noise, and feature heterogeneity, which require careful preprocessing for effective anomaly detection.

\subsection{Data Preprocessing}
To evaluate the effectiveness of LLMs in enhancing anomaly detection, two distinct preprocessing strategies are employed:

\subsubsection{Traditional Preprocessing Using PCA}
In the traditional preprocessing approach using Principal Component Analysis (PCA), the steps included standardizing numerical features through Min-Max scaling and one-hot encoding of categorical features such as $protocol\_type$ and $flag$. PCA is then applied to reduce the input feature space to 25 principal components, preserving approximately 95\% of the variance. The resulting PCA-transformed data served as input to the anomaly detection model.

\subsubsection{LLM\textminus Assisted Preprocessing}
The second approach leveraged GPT\textminus4 to automate and optimize the preprocessing pipeline. The LLM analyzed the dataset by reviewing feature statistics, such as variance and correlation, to identify redundant and low\textminus variance features. It recommended specific feature transformations, such as binarization for sparse columns and averaging for highly correlated features, and suggested encoding techniques for categorical variables to ensure compatibility with machine learning models. By generating automated preprocessing scripts, the LLM\textminus assisted method reduced manual effort and produced cleaner, more meaningful input data for anomaly detection.

\subsection{Anomaly Detection Using Autoencoder}
The preprocessed datasets from both methods (PCA-based and LLM-assisted) are used to train separate Autoencoder models. An Autoencoder is a type of unsupervised neural network that learns to compress and reconstruct input data. During inference, instances with high reconstruction errors are flagged as anomalies.

The Autoencoder model consists of two main components:
\begin{align}
    z &= \sigma(W_e x + b_e), \quad z \in \mathbb{R}^m, \; m < n \label{eq:encoder} \\
    \hat{x} &= \sigma(W_d z + b_d), \quad \hat{x} \in \mathbb{R}^n \label{eq:decoder}
\end{align}

where \( x \in \mathbb{R}^n \) represents the input data, \( z \) is the latent representation, and \( \hat{x} \) is the reconstructed input. The parameters \( W_e, W_d \) and biases \( b_e, b_d \) are learned during training. The reconstruction loss is computed as:
\begin{equation}
    L_{\text{recon}} = \frac{1}{n} \sum_{i=1}^n (x_i - \hat{x}_i)^2
\end{equation}
Anomalies are detected by thresholding the reconstruction error.

\subsection{LLM-Driven Anomaly Explanation}
To bridge the interpretability gap in anomaly detection, GPT-4 is integrated into the framework to generate human-readable explanations for detected anomalies. The process involves:
\begin{enumerate}
    \item Extracting anomaly-related features, such as reconstruction error, \textit{src\_bytes}, and \textit{protocol\_type}.
    \item Structuring the data into natural language prompts and feeding them to GPT-4.
    \item Generating textual explanations that describe the anomaly\textquotesingle
s potential cause and implications.
\end{enumerate}
For example, an anomaly with unusually low traffic volume (\textit{src\_bytes}) and a rejected TCP connection may be explained as a failed network scanning attempt.


\subsection{Comparison and Evaluation}
The comparison and evaluation of the two preprocessing approaches were conducted using multiple metrics to assess the overall performance of the anomaly detection framework. The evaluation included the reconstruction error distributions of the Autoencoder, as well as training and validation loss curves, to analyze the model's convergence and generalization capabilities. Additionally, standard anomaly detection performance metrics, such as accuracy, false positive rate, and macro-average precision, recall, and F1-score for both the normal and attack classes, were used to provide a comprehensive assessment. The macro-average F1-score, in particular, highlighted the balance between precision and recall for the two classes, ensuring the model's ability to detect rare anomalies without bias toward the dominant class.

Furthermore, the quality of GPT-4 explanations was evaluated in terms of their ability to generate actionable insights for detected anomalies, enhancing the interpretability of the system. The comparative results clearly demonstrated that the LLM-assisted preprocessing approach not only improved anomaly detection accuracy but also provided more meaningful and interpretable outputs, making the system more reliable for real-world applications.

\begin{figure}[!t]
    \centering
    \includegraphics[width=0.45\textwidth]{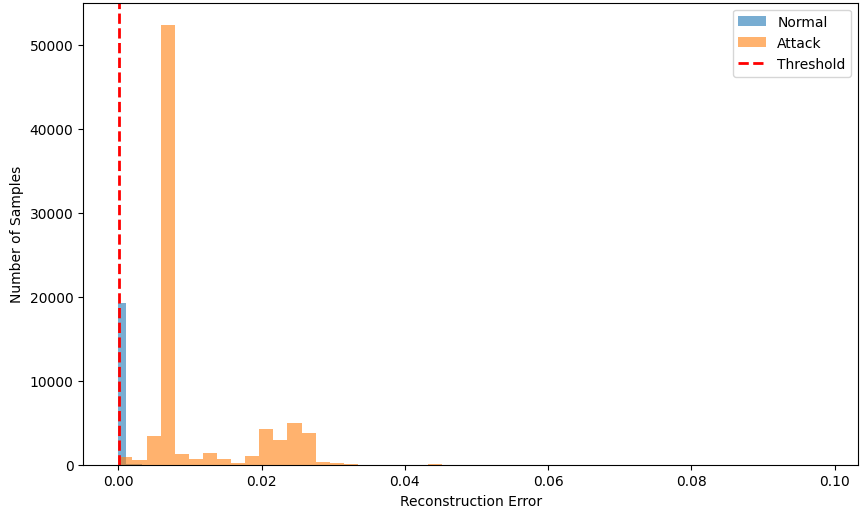}
    \caption{Reconstruction error distribution of the LLM-integrated Autoencoder.}
    \label{fig:llm_recon}
\end{figure}

\section{Results and Discussion}

This section presents the results of our proposed framework, comparing the performance of traditional numerical anomaly detection methods (PCA-based preprocessing) with LLM-assisted preprocessing. Additionally, we demonstrate the ability of LLMs to provide interpretable explanations for detected anomalies. The results are analyzed in terms of training/validation loss, reconstruction error distribution, and performance metrics.



\subsection{Autoencoder Performance Analysis}

To evaluate the impact of LLM-assisted preprocessing, two Autoencoder models are trained:
\begin{itemize}
    \item \textbf{Traditional Autoencoder}: Using PCA-based preprocessing with 25 principal components.
    \item \textbf{LLM insights Integrated Autoencoder}: Using LLM-driven preprocessing, where GPT optimized data preparation by suggesting feature transformations, removal of redundancy, and encoding strategies.
\end{itemize}



\subsubsection{Reconstruction Error Distribution}

\begin{figure}[!t]
    \centering
    \includegraphics[width=0.45\textwidth]{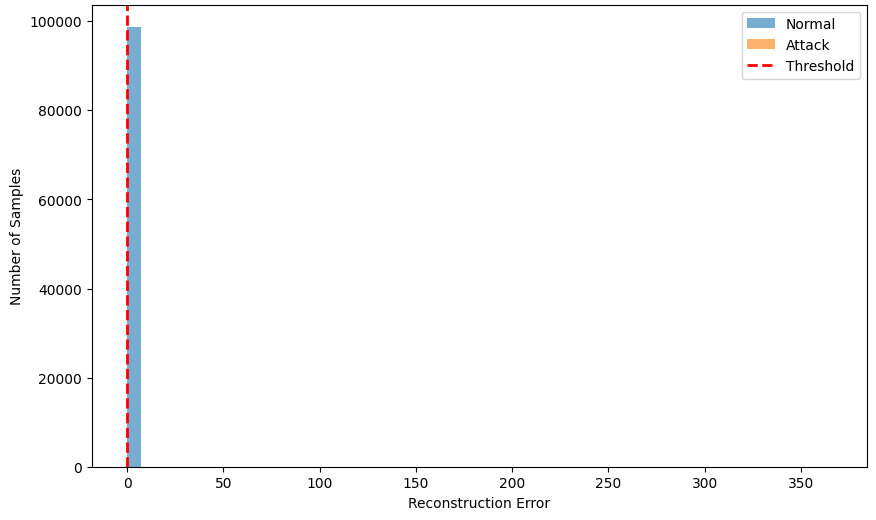}
    \caption{Reconstruction error distribution of the traditional Autoencoder.}
    \label{fig:traditional_recon}
\end{figure}

The reconstruction error distributions for both models are compared in Fig.~\ref{fig:traditional_recon} and Fig.~\ref{fig:llm_recon}. The LLM-integrated Autoencoder produces a tighter error distribution for normal samples and more distinct outliers for anomalies. This indicates that the LLM-assisted preprocessing enhanced the model's ability to differentiate between normal and anomalous patterns.

\subsubsection{Performance Comparison}

\begin{figure}[!t]
    \centering
    \includegraphics[width=0.30\textwidth]{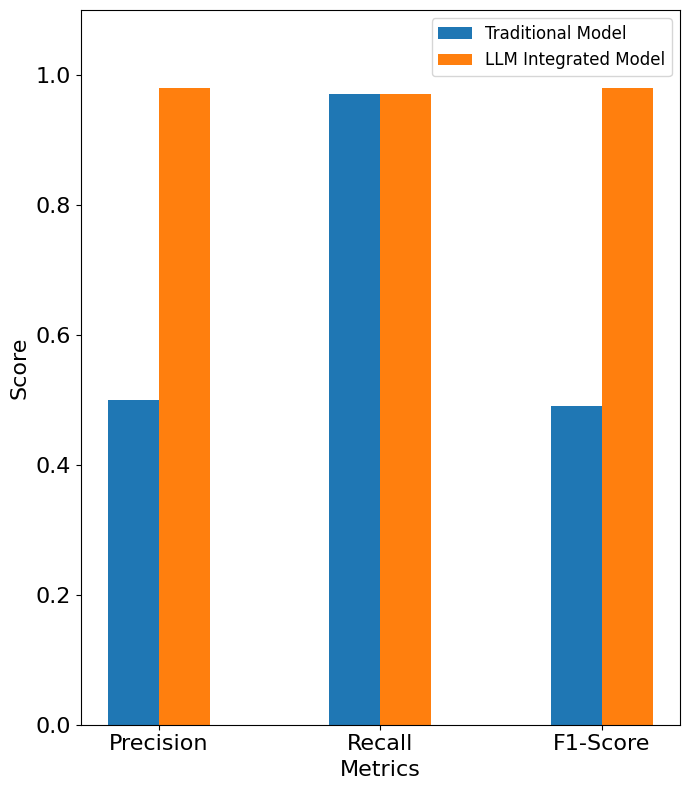}
    \caption{Comparison of macro-average performance metrics between the traditional Autoencoder and the LLM-integrated Autoencoder.}
    \label{fig:performance_comparison}
\end{figure}

Fig.~\ref{fig:performance_comparison} provides a performance comparison between the traditional Autoencoder and the LLM-integrated Autoencoder. The LLM-integrated model achieves a higher accuracy and lower false positive rate due to its improved preprocessing pipeline.

\subsection{LLM-Driven Explanation of Anomalies}

The integration of GPT-4 allows for human-readable explanations of detected anomalies. This section demonstrates the ability of the LLM to analyze anomalies flagged by the Autoencoder and provide actionable insights for cybersecurity analysts.

Figs.~\ref{fig:anomaly_explanation1} and \ref{fig:anomaly_explanation2} illustrate example anomalies detected by the Autoencoder, along with their GPT-generated explanations. These explanations provide contextual insights, such as potential attack vectors, failed connection attempts, or unusual traffic patterns.

The ability of GPT-4 to describe anomalies in natural language enhances the interpretability of the system. For instance, anomalies with high reconstruction errors may be explained as unusual TCP connection refusals or potential port scanning activities. Such insights are crucial for cybersecurity analysts to take informed actions.

\subsection{Discussion}

The experimental results clearly demonstrate the advantages of the proposed framework. First, improved detection performance was achieved through LLM-assisted preprocessing, which refined the data input and led to more accurate anomaly detection. Second, the faster convergence of the LLM-integrated Autoencoder was evident, as it required fewer epochs to converge, as shown in the loss curves. Third, the framework provided enhanced interpretability by leveraging GPT-4 to generate actionable insights for anomalies, effectively bridging the gap between numerical outputs and human understanding. By combining numerical AI models with advanced language capabilities, the proposed framework addresses the limitations of existing anomaly detection systems and fosters trust in AI-driven cybersecurity for critical infrastructure.

\begin{figure}[!t]
    \centering
    \begin{tcolorbox}[title=Prompt Response: Anomaly 1, colframe=black, colback=white, sharp corners=south]
    \small
    \textbf{Reconstruction Error:} 0.0001

    \textbf{Features:}
    \begin{itemize}
        \item \texttt{src\_bytes=1.4883707192251517e-06}, \texttt{dst\_bytes=0.0}
        \item \texttt{protocol\_type\_tcp=0.0}, \texttt{flag\_REJ=0.0}
    \end{itemize}

    \textbf{Prediction:} Attack

    \textbf{Generated Insight:}  
    The anomaly suggests minimal source bytes (\texttt{src\_bytes}) and zero destination bytes (\texttt{dst\_bytes}). The TCP protocol is inactive (\texttt{protocol\_type\_tcp=0.0}), and there are no rejection flags (\texttt{flag\_REJ=0.0}). This behavior might indicate \textbf{low-volume traffic anomalies or stealth reconnaissance} attempts where connections do not exchange meaningful data.

    \textbf{Steps for Further Analysis:}
    \begin{enumerate}
        \item \textbf{Network Logs:} Verify the source IP and check for repeated connection attempts.
        \item \textbf{Port Analysis:} Determine if specific ports are targeted.
        \item \textbf{Threshold Adjustment:} Reassess detection thresholds for anomalies with near-zero payloads.
    \end{enumerate}
    \end{tcolorbox}
    \caption{Example 1: GPT-generated explanation of a detected anomaly.}
    \label{fig:anomaly_explanation1}
\end{figure}

\begin{figure}[ht]
    \centering
    \begin{tcolorbox}[title=Prompt Response: Anomaly 2, colframe=black, colback=white, sharp corners=south]
    \small
    \textbf{Reconstruction Error:} 0.0008

    \textbf{Features:}
    \begin{itemize}
        \item \texttt{src\_bytes=3.9661041452220616e-07}, \texttt{dst\_bytes=6.013033152373364e-05}
        \item \texttt{protocol\_type\_tcp=1.0}, \texttt{flag\_REJ=0.0}
    \end{itemize}

    \textbf{Prediction:} Attack

    \textbf{Generated Insight:}  
    The anomaly indicates a very small \texttt{src\_bytes} value and slightly higher \texttt{dst\_bytes}, while the TCP protocol is active (\texttt{protocol\_type\_tcp=1.0}). There are no rejections (\texttt{flag\_REJ=0.0}). This could indicate \textbf{port scanning or probing behavior} where minimal traffic is sent to test open ports or system responses.

    \textbf{Steps for Further Analysis:}
    \begin{enumerate}
        \item \textbf{Traffic Context:} Analyze surrounding traffic for sequential port scans.
        \item \textbf{Protocol Analysis:} Verify if the TCP handshake completes or drops midway.
        \item \textbf{Source IP Tracking:} Look for repeated connections from the same IP.
    \end{enumerate}
    \end{tcolorbox}
    \caption{Example 2: GPT-generated explanation of a detected anomaly.}
    \label{fig:anomaly_explanation2}
\end{figure}

\section{Conclusion}
This paper proposed a hybrid framework for anomaly detection in IoT systems, integrating Autoencoder-based models with GPT-4 for enhanced preprocessing and interpretability. Two approaches were explored: a traditional PCA-based method and an LLM-assisted pipeline where GPT-4 dynamically optimized feature selection and transformation.

Experimental results on the KDDCup99 dataset showed that the LLM-assisted preprocessing improved anomaly detection performance, leading to faster convergence, better reconstruction error separation, and enhanced accuracy. Additionally, GPT-4 provided natural language explanations for detected anomalies, improving interpretability and aiding human decision-making.

In summary, the proposed framework demonstrates the potential of combining numerical AI models with LLMs to achieve accurate and explainable anomaly detection in IoT systems. Future work will focus on real-time IoT log integration and validation across additional critical infrastructure datasets.

\ifCLASSOPTIONcaptionsoff
  \newpage
\fi

\bibliographystyle{IEEEtran}

\begin{thebibliography}{10}
\providecommand{\url}[1]{#1}
\csname url@rmstyle\endcsname
\providecommand{\newblock}{\relax}
\providecommand{\bibinfo}[2]{#2}
\providecommand\BIBentrySTDinterwordspacing{\spaceskip=0pt\relax}
\providecommand\BIBentryALTinterwordstretchfactor{4}
\providecommand\BIBentryALTinterwordspacing{\spaceskip=\fontdimen2\font plus
\BIBentryALTinterwordstretchfactor\fontdimen3\font minus
  \fontdimen4\font\relax}
\providecommand\BIBforeignlanguage[2]{{%
\expandafter\ifx\csname l@#1\endcsname\relax
\typeout{** WARNING: IEEEtran.bst: No hyphenation pattern has been}%
\typeout{** loaded for the language `#1'. Using the pattern for}%
\typeout{** the default language instead.}%
\else
\language=\csname l@#1\endcsname
\fi
#2}}

\bibitem{djenna2021internet}
A.~Djenna, S.~Harous, and D.~E. Saidouni, ``Internet of things meet internet of
  threats: New concern cyber security issues of critical cyber
  infrastructure,'' \emph{Applied Sciences}, vol.~11, no.~10, p. 4580, 2021.

\bibitem{aslan2023comprehensive}
{\"O}.~Aslan, S.~S. Aktu{\u{g}}, M.~Ozkan-Okay, A.~A. Yilmaz, and E.~Akin, ``A
  comprehensive review of cyber security vulnerabilities, threats, attacks, and
  solutions,'' \emph{Electronics}, vol.~12, no.~6, p. 1333, 2023.

\bibitem{chandola2009anomaly}
V.~Chandola, A.~Banerjee, and V.~Kumar, ``Anomaly detection: A survey,''
  \emph{ACM computing surveys (CSUR)}, vol.~41, no.~3, pp. 1--58, 2009.

\bibitem{su2024large}
J.~Su, C.~Jiang, X.~Jin, Y.~Qiao, T.~Xiao, H.~Ma, R.~Wei, Z.~Jing, J.~Xu, and
  J.~Lin, ``Large language models for forecasting and anomaly detection: A
  systematic literature review,'' \emph{arXiv preprint arXiv:2402.10350}, 2024.

\bibitem{liu2019secure}
X.~Liu, C.~Qian, W.~G. Hatcher, H.~Xu, W.~Liao, and W.~Yu, ``Secure internet of
  things (iot)-based smart-world critical infrastructures: Survey, case study
  and research opportunities,'' \emph{IEEE Access}, vol.~7, pp.
  79\,523--79\,544, 2019.

\bibitem{ghimire2023implementation}
A.~Ghimire, A.~N. Asiri, B.~Hildebrand, and F.~Amsaad, ``Implementation of
  secure and privacy-aware ai hardware using distributed federated learning,''
  in \emph{2023 IEEE 16th Dallas Circuits and Systems Conference (DCAS)}.\hskip
  1em plus 0.5em minus 0.4em\relax IEEE, 2023, pp. 1--6.

\bibitem{ravipati2019intrusion}
R.~D. Ravipati and M.~Abualkibash, ``Intrusion detection system classification
  using different machine learning algorithms on kdd-99 and nsl-kdd datasets-a
  review paper,'' \emph{International Journal of Computer Science \&
  Information Technology (IJCSIT) Vol}, vol.~11, 2019.

\bibitem{gao2022tsmae}
H.~Gao, B.~Qiu, R.~J.~D. Barroso, W.~Hussain, Y.~Xu, and X.~Wang, ``Tsmae: a
  novel anomaly detection approach for internet of things time series data
  using memory-augmented autoencoder,'' \emph{IEEE Transactions on network
  science and engineering}, vol.~10, no.~5, pp. 2978--2990, 2022.

\bibitem{bertero2017experience}
C.~Bertero, M.~Roy, C.~Sauvanaud, and G.~Tr{\'e}dan, ``Experience report: Log
  mining using natural language processing and application to anomaly
  detection,'' in \emph{2017 IEEE 28th International Symposium on Software
  Reliability Engineering (ISSRE)}.\hskip 1em plus 0.5em minus 0.4em\relax
  IEEE, 2017, pp. 351--360.

\bibitem{hadadi2024anomaly}
F.~Hadadi, Q.~Xu, D.~Bianculli, and L.~Briand, ``Anomaly detection on unstable
  logs with gpt models,'' \emph{arXiv preprint arXiv:2406.07467}, 2024.

\bibitem{o2023deployment}
J.~O'Brien, S.~Ee, and Z.~Williams, ``Deployment corrections: An incident
  response framework for frontier ai models,'' \emph{arXiv preprint
  arXiv:2310.00328}, 2023.

\bibitem{wang2017heterogeneous}
L.~Wang, ``Heterogeneous data and big data analytics,'' \emph{Automatic Control
  and Information Sciences}, vol.~3, no.~1, pp. 8--15, 2017.

\bibitem{tavallaee2009detailed}
M.~Tavallaee, E.~Bagheri, W.~Lu, and A.~A. Ghorbani, ``A detailed analysis of
  the kdd cup 99 data set,'' in \emph{2009 IEEE symposium on computational
  intelligence for security and defense applications}.\hskip 1em plus 0.5em
  minus 0.4em\relax Ieee, 2009, pp. 1--6.

\end{thebibliography}

\vfill


\end{document}